\documentclass[DIV=15,
                  twocolumn,
               ]{scrartcl}  
\usepackage{graphicx}
\usepackage{amsmath}
\usepackage{xcolor}
\usepackage{hyperref}
\usepackage[square, numbers, sort&compress]{natbib}   
\usepackage{color}
\usepackage{lipsum}
\usepackage{multicol}

\usepackage{soul}
\usepackage{subfig}
\usepackage{authblk} 

\usepackage{fancyhdr}
\fancypagestyle{plain}{%
  \fancyhf{}%
  \fancyfoot[L]{
  \vspace{-1.25cm} 
  \footnotesize{this corresponds to a preprint \textit{\journal{}}  \hfill
  }} }

\begin{document}

\title{Accurate prediction of melt pool shapes in laser powder bed fusion by the non-linear temperature equation including phase changes
}
\subtitle{Model validity: isotropic versus anisotropic conductivity} 


\author[1]{Stefan Kollmannsberger}
\author[2]{Massimo Carraturo}
\author[2]{Alessandro Reali}
\author[2]{Ferdinando Auricchio}

\affil[1]{Chair for Computation in Engineering,
Technical University of Munich,
Arcisstr. 21, 80333 M\"unchen, Germany}

\affil[2]{University of Pavia, Italy}

\newcommand{\journal}{submitted to IMMI: Integrating Materials and Manufacturing Innovation, Springer}
\newcommand{\publicationDate}{\today}


\maketitle

\subsection*{Abstract}
In this contribution, we validate a physical model based on a transient temperature equation (including latent heat) w.r.t. the experimental set AMB2018-02 provided within the additive manufacturing benchmark series, established at the National Institute of Standards and Technology, USA. 
We aim at predicting the following quantities of interest: width, depth, and length of the melt pool by numerical simulation and report also on the obtainable numerical results of the cooling rate.
\\
We first assume the laser to posses a double ellipsoidal shape and demonstrate that a well calibrated, purely thermal model based on isotropic thermal conductivity is able to predict all the quantities of interest, up to a deviation of maximum 7.3\% from the experimentally measured values.
However, it is interesting to observe that if we directly introduce, whenever available, the measured laser profile in the model (instead of the double ellipsoidal shape) the investigated model returns a deviation of 19.3\% from the experimental values.
This motivates a model update by introducing anisotropic conductivity, which is intended to be a simplistic model for heat material convection inside the melt pool.
Such an anisotropic model enables the prediction of all quantities of interest mentioned above with a maximum deviation from the experimental values of 6.5\%.
%
We note that, although more predictive, the anisotropic model induces only a marginal increase in computational complexity. 

\section{Introduction}
\label{intro}
The shape and thermal history of the melt pool are key ingredient to determine the physical properties of an artifact generated through a welding process. 
Therefore, the prediction of weld pool dynamics has been a subject of intensive research in the last decades in both the experimental and the numerical modeling community of welding;
e.g., recent reviews of this subject are provided in~\cite{Fotovvati2018,Svenungsson2015}.
Furthermore, it is fundamental to observe that process-structure-property relationships are also tightly interlinked and strongly determined by the characteristics of the weld pool in laser powder bed fusion (LPBF) additive manufacturing technologies~\cite{Smith2016a}. 
Therefore, an accurate thermal analysis is a key ingredient in the numerical simulations and predictions of LPBF processes as well.
\par
To this end, many physical models have been proposed to obtain accurate and reliable numerical approximations of melt pools. 
Although different in scale, the basic phenomena in LPBF are similar to those in arc welding processes, see~\cite{Tanaka2014,Lindgren2007,Goldak2005} for an overview. 
Recent summaries more specific to LPBF processes are published in~\cite{Megahed2016,Smith2016a,Schoinochoritis2017}. 
While particle based models~\citep{Khairallah2016} as well as Lattice Boltzmann type approaches~\citep{Korner2011} exist, most common are continuum models based on the conservation of mass, momentum and energy~\citep{Yan2018a}. 
Continuum approaches allow for modeling the transient evolution of primal variables (temperatures, pressures, and velocities) taking into account a large number of effects, such as the convection inside the melt pool, also including the one caused by a gradient in the surface tension (Marangoni effect as well as capillary effects), vaporization, momentum losses in mushy zones due to porous media effects, etc.
\par
All these models may deliver very accurate results, but the more effects they include, the more computational power they require. 
Additionally, the abundance of models comes along with a wealth of parameters: these may be material viscosity, density, thermal conductivity and capacity, latent heat, etc., most of which show a non-negligible temperature dependence, such that their accurate, experimental determination may be both crucial and critical. 
Further modeling parameters, such as emissivity or absorptivity or even the geometry of powder particles, may come into play and they can be introduced in the model as boundary or initial conditions. 
However, accurate measurements of many of the listed parameters are not publicly available. 
This is even true for the most basic parameters, such as heat capacity or thermal conductivity, that are not published for the temperatures involved in metal based LPBF processes. 
All this drought of information results in the fact that even the evaluation of several parameters may itself often rely on models which, in turn, need to be calibrated against further observations.

The dilemma of choosing a correct model for the case of limited data is an important issue in statistics. 
As an example, George Box~\cite{Box1976} stated, somewhat drastically in his well known aphorism, that all models are wrong and that, therefore, the most complicated model is not necessarily the best. 
Instead, it is recommended to follow the lines of William of Occam, in which an economical description of the observations is sought which `is as simple as possible, but not simpler'. 

Following this line of thought, the purpose of the present paper is to built and validate an economical model able to replicate the results obtained by the benchmark measurements of a single line laser stroke on a bear metal plate of IN 625 published in~\cite{ambench2018}. 
As stated in the chapter CHAL-AMB2018-02-MP of the previous reference, the quantities of interest are the width, depth, and length of the melt pool. Additionally, we also monitor the cooling rate as defined in CHAL-AMB2018-02-CR, even if this quantity is not of primary concern here. 
To this end, we employ a heat transfer model which considers the different phases of the material as a homogeneous media. This model is designed to work for pure conduction, i.e. keyhole modes can neither be forecast nor replicated.
This approach is well established in literature and has proven to be effective also in the thermal numerical analysis of large scale LPBF processes~\cite{Chiumenti2017a,Riedlbauer2017,Denlinger2016}.
Other successful attempts in this direction, however with a focus on the scale of the melt pool, include the very recent publication of Zhang~\cite{Zhang2019}, which provides a summary of previous approaches but most importantly also incorporates anisotropic conductivities, as discussed in the paper at hand. 
The model proposed in~\cite{Zhang2019} is more elaborate than ours as it also incorporates a spatially variable laser absorptivity and, in this sense, it is not minimal w.r.t. the data set we face. 
Further, interesting efforts to construct a valid, yet minimal model, are published in~\cite{Mahmoudi2018} where the linking of thermal models to experiments is carried out via surrogate modeling based on multivariate Gaussian processes.

We begin by introducing the widely used physical model based on the transient heat equation including phase changes in~\autoref{sec:governingEquations}. 
We shortly remark on the verification of this model in~\autoref{sec:ModelVerification}, before we move to model validation in~\autoref{sec:ModelValidation}. 
The section on model validation is the main section and commences with reciting the main results of the benchmark cases obtained on two machines, a commercial machine (CBM) and the additive manufacturing metrology testbed (AMMT), both located at the National Institute of Standards and Technology, USA. 
In an effort to obtain a minimal set of modeling parameters, we evaluate the sensitivities of the quantities of interest to the modeling parameters given by the physical model. 
We then select only the most relevant modeling parameters and use them to calibrate the physical model towards a similar benchmark already published as case 7 in~\cite{Ghosh2018}. We then proceed with the evaluation of our model against the benchmark results on the CBM machine. 
We observe that the more accurate measurements of the laser profile on the AMMT render the model calibrated to the CBM machine using a double-ellipsoidal heat source less accurate in the AMMT case in which accurate measurements of the laser profile exist. 
This observation necessitates an update of the model. 
The model update is presented in~\autoref{sec:anisotrConductivity} by incorporating anisotropic conductivity, which is thought to be a simple way to model the convection inside the melt pool. Finally, in~\autoref{sec:Conclusions} we conclude that given accurate measurements of the profile of the laser, the anisotropic model provides an increase in accuracy over the tested parameter range as compared to the simpler, isotropic physical model.

\section{Governing equations} \label{sec:governingEquations}

We use a non-linear heat transfer equation as a physical model to describe the evolution of temperature $T=T(t,\mathbf{x})$ as a function of space and time. 
Given a spatial domain $\Omega$ and a time interval $\mathcal{T}=\left[0,t_{end}\right)$, the heat transfer equation can be written as follows:
\begin{equation}
\rho c \frac{\partial T}{\partial t} +
\rho L \frac{\partial f_{pc}}{\partial t}-
\nabla \cdot\left( k \nabla T \right) = 0
\hspace{1cm} \text{in}  \,\quad \Omega\times\mathcal{T}.
\label{eq:transientHeat}
\end{equation}
Therein $\rho$
and
$L$ describe the density and the latent heat of the material,
$c=c(T,\mathbf{x})$ and
$k=k(T,\mathbf{x})$ are the temperature dependent heat capacity and thermal conductivity of the material,
while $f_{pc}=f_{pc}(T)$ is the phase-change function describing the solid-to-liquid phase transition of the material.
Therefore, beside the non linear contribution of the heat capacity and thermal conductivity, the latent heat term of~\autoref{eq:transientHeat} introduces a further nonlinearity into the problem.
\par
\autoref{eq:transientHeat} is completed by the initial condition at time $t=0$:
\begin{equation}
T(\mathbf{x},0)=T_0\hspace{1cm} \text{in}  \,\quad \Omega,
\label{eq:InitialConditions}
\end{equation}
as well as Neumann boundary conditions:
\begin{equation}
k \nabla T \cdot \mathbf{n} = q^r + q^l\hspace{1cm} \text{on} \,\quad \Gamma_N\times\mathcal{T}.
\label{eq:emissivity}
\end{equation}
Herein, $T_0$ is the initial temperature of the body,
$\mathbf{n}$ is the unit normal vector, 
$q^l$ is the heat flux input 
%
and $q^r$ is the radiation boundary condition defined as:
\begin{equation}
q^r=\sigma \epsilon \left(T^2+T^2_{e} \right) \left(T^2_{e} - T^2 \right). 
\label{eq:RadiationBoundaryCondition}
\end{equation}
In \autoref{eq:RadiationBoundaryCondition}
$\sigma$ is the Stefan-Boltzmann constant, 
$\epsilon$ is the emissivity of the material, 
and $T_{e}$ is the ambient temperature.
In our model, convection boundary conditions are neglected. 
Further details, specifically the adopted finite element formulation, are provided in~\cite{Celentano1994,Kollmannsberger2018}. 
\subsection{Phase-change model}
\begin{figure}
\centering 
\includegraphics[width=0.4\textwidth]{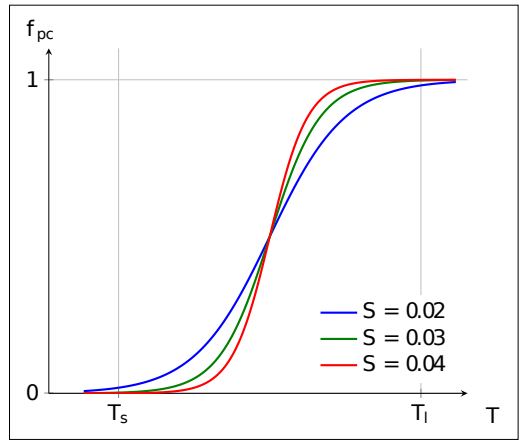}
\caption{Phase change function for different values of the parameter $S$\label{pcf}}
\end{figure}

For iso-thermal phase changes $f_{pc}$ exhibits a jump at the melting temperature $T_m$, as the temperature changes the material state from solid to liquid.
Since the phase-change for metals is actually non-isothermal, we regularize this sudden jump between two temperatures $T_s$ and $T_l$, with $T_s<T_l$. We can now define the phase change function $f_{pc}$, such as:
\begin{equation}
f_{pc}(T)=\frac{1}{2}\left[ \left(S\frac{2}{T_l-T_s} \left(T-\frac{T_s+T_l}{2} \right)  \right)+1 \right].
\label{eq:phaseChangeFunction}
\end{equation}
The parameter $S$ in~\autoref{eq:phaseChangeFunction} is initially estimated such that the bulk of the phase change occurs between $T_s$ and $T_l$ (see Figure~\ref{pcf}). Nevertheless, since no measurement data is available as to how exactly the phase change occurs, $S$ also requires calibration.

\subsection{Heat flux model}\label{sec:heatFluxModel}
\begin{figure}[h!]
\centering
     \includegraphics[width=0.5\textwidth]{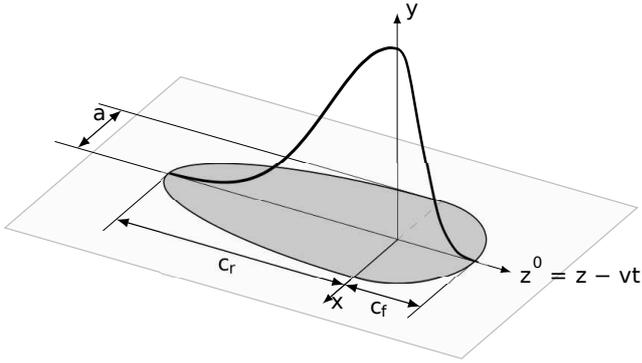}
      \caption{Goldak model for the heat flux input. The model consists in a double-ellipse on which a gaussian profile is defined.\label{GaussDist}}
\end{figure}

In the sequel we consider two variants of the heat flux input $q^l$. 
The first variant is the double elliptical model of Goldak~\cite{Goldak1984} described in~\autoref{GaussDist}. The front quadrant as is defined by:
\begin{equation}
q^l=\dfrac{2\text{Q}\eta f_{f}}{\pi a c_f}\text{exp}\left(-2((z^{'}-z^{'}_0)/c_f^2+(x-x_0)/a^2)\right),
\label{eq:GaussFrontf}
\end{equation}
while in the rear quadrant it takes the form:
\begin{equation}
q^l=\dfrac{2\text{Q}\eta f_{r}}{\pi a c_r}\text{exp}\left(-2((z^{'}-z^{'}_0)/c_r^2+(x-x_0)/a^2)\right).
\label{eq:GaussFrontr}
\end{equation}
Herein, $Q$ is the laser power and 
$\eta$ is the absorptivity of the material. 
The geometrical parameters $z^{'}_0$ 
and $x_0$ define the center of the laser beam on the upper surface at time $t$, while
$f_{f}$ and $f_{r}$ are the fraction of heat deposited in the front and the rear quadrant respectively, which have the side condition that $f_r+f_f=2$ (see~\cite{Goldak1984} for further details).

The second variant the heat source $q^l$ is not a model. In fact, $q^l$ is directly provided to~\autoref{eq:emissivity} as given by corresponding measurements. 
\section{Model Verification} \label{sec:ModelVerification}
The computational model was verified against the series of analytical or semi-analytical solutions defined in~\cite{Kollmannsberger2018}, where a multi-level $hp$-discretization was used. 
The computational model utilized in the paper at hand is slightly different, as it uses an IGA discretization wherein multi-level B{\'e}zier extraction is applied to construct an efficient discretization which is refined locally in the vicinity of the laser beam. 
This implementation was verified in~\cite{DAngella2018a} in two dimensions as well as in three dimensions using the same series of problems which were also used for the verification of the multi-level $hp$-basis~\cite{Kollmannsberger2018}. 
Since the focus of the present contribution is the validation of the model, we will not repeat these extensive verification studies. 
Instead, in the next section of this paper, we will use the capabilities of the proposed discretization to directly evaluate the validity of the physical model given in~\autoref{sec:governingEquations}. 

\section{Model Validation} \label{sec:ModelValidation}
As a preamble to this section we want to highlight the fact that there are situations (e.g. the presence of highly complex phenomena, problem physics still unclear, model uncertainties and difficulties in ascertain its effectiveness, inability to measure all the model parameters) in which model validation must consist of two steps. 
In the first step (calibration step) the indeterminacy of the physical model is investigated and calibrated against a first set of experimental evidences; 
in the second step (validation step) the numerical results are compared against a different set of experimental evidences in order to define the range of validity and the robustness of the numerical model.
The case under investigation is characterized by the inability to measure  all the model parameters, in particular we have limited information on the absorptivity, emissivity, thermal conductivity and heat capacity of the material at high temperature,  justifying the choice of the previously defined two-step model validation process.
%
\par
In~\autoref{sec:BenchmarkCases} we will shortly describe and report the experimental benchmarks published in~\cite{ambench2018}. 
Following the previously described steps of the validation procedure, first, in~\autoref{sec:ModelCalibrationStandardHeatSource}, we  calibrate the isotropic model of~\autoref{sec:governingEquations} using the double ellipsoidal heat source introduced by Goldak~\cite{Goldak1984} and then we validate the isotropic model using the two heat fluxes described in~\autoref{sec:heatFluxModel}. 
For the case in which an accurate measurement of the laser power distribution is given, we observe that the isotropic material assumption has a very limited range of validity.  
These findings serve as a motivation to extend the physical model by introducing anisotropic conductivities. 
This extended model is then presented in~\autoref{sec:anisotrConductivity} where it will be demonstrated that it predicts weld pool shapes with an improved accuracy.
\subsection{Benchmark cases \label{sec:BenchmarkCases}}
All benchmark cases are thoroughly defined in the laser additive manufacturing benchmarks published in~\cite{ambench2018}, including a detailed report on the measurements. 
The benchmarks are obtained through a traveling laser beam on a bear metal plate of nickel-based alloy IN625. 
The experimental quantities we will use to validate our model are: width, length, and depth of the melt pool as defined in CHAL-AMB2018-02-MP of the above reference. 
We also report on the cooling rates defined in CHAL-AMB2018-02-CR, although they are not the primal focus in the paper at hand. 
The benchmarks cited above and reported in~\cite{ambench2018} were performed on two different machines: a commercial machine (CBM) and the additive manufacturing metrology testbed (AMMT), both located at the National Institute of Standards and Technology, USA. 
On each machine a set of ten measurements was carried out for three different cases (labeled A, B, and C), i.e. for varying laser power and speed. These cases are specified in~\autoref{tab:CBSparameterValues} for the CBM machine and~\autoref{tab:AMMTparameterValues} for the AMMT machine. 
The averages of the experimental measurements for the CBM machine are reported in~\autoref{tab:CBSMeasurements}, while average measurements for the AMMT machine are reported in~\autoref{tab:AMMTMeasurements}. In the first case the cooling rate is defined as:
\begin{equation*}
CR=\dfrac{1290\left[^{\circ}\text{C}\right]-1000\left[^{\circ}\text{C}\right]}{\Delta d [mm]}\times v \left[\dfrac{mm}{sec}\right],
\end{equation*}
with $v$ laser speed and $\Delta d$ distance in the direction of the laser path, while in the second case as:
\begin{equation*}
CR=\dfrac{1290\left[^{\circ}\text{C}\right]-1190\left[^{\circ}\text{C}\right]}{\Delta d [mm]}\times v \left[\dfrac{mm}{sec}\right].
\end{equation*}
For the exact definition of $v$ and $\Delta d$, as well as for further details on the experimental benchmarks we refer to the original website which continues to be updated as further measurements become available~\cite{ambench2018}. 
%

\begin{table}
\centering
\begin{tabular}{llll}
\hline\noalign{\smallskip}
Parameter values              & A      & B      & C  \\
\noalign{\smallskip}\hline\noalign{\smallskip}
laser power [W]               & 150 & 195   & 195 \\
laser speed [mm/s]            & 400   & 800     & 1200 \\
laser spot diameter $D4\sigma [\mu m]$ & 100 & 100   & 100 \\
\noalign{\smallskip}\hline
\end{tabular}
\caption{CBM machine: parameter values \label{tab:CBSparameterValues}}
\end{table}

\begin{table}
\centering
\begin{tabular}{llll}
\hline\noalign{\smallskip}
Parameter values              & A      & B      & C  \\
\noalign{\smallskip}\hline\noalign{\smallskip}
laser power [W]               & 137.9 & 179.2   & 179.2 \\
laser speed [mm/s]            & 400   & 800     & 1200 \\
laser spot diameter $D4\sigma [\mu m]$ & 170 & 170   & 170 \\
\noalign{\smallskip}\hline
\end{tabular}
\caption{AMMT machine: parameter values \label{tab:AMMTparameterValues}}
\end{table}

\begin{table}
\centering
\begin{tabular}{lll}
\hline\noalign{\smallskip}
case                  & length    &    cooling rate             \\
                      & $[\mu m]$ &      [$\frac{^\circ\text{C}}{sec}$]   \\
\noalign{\smallskip}\hline\noalign{\smallskip}
A            & 659 $\pm$ 21 &$6.20\times 10^5\pm 7.99\times 10^4$ \\
B             & 782 $\pm$ 21 &        $9.35\times 10^5\pm 1.43\times 10^5$ \\
C            & 754  $\pm$ 46 & $1.28\times 10^6\pm 3.94\times 10^5$ \\
\noalign{\smallskip}\hline
\end{tabular}
\caption{CBM machine: experimental measurements according to~\cite{ambench2018}, CHAL-AMB2018-02-MP~ \label{tab:CBSMeasurements}}
\end{table}

\begin{table}
\centering
\begin{tabular}{lllll}
\hline\noalign{\smallskip}
case                  & length    & width            & depth             & cooling rate             \\
                      & $[\mu m]$ &  $[\mu m]$       &  $[\mu m]$        &  [$\frac{^\circ\text{C}}{sec}$]   \\
\noalign{\smallskip}\hline\noalign{\smallskip}
A            & 300  & 147.9   & 42.5 & $1.16\times 10^6$ \\
B             & 359 & 123.5   & 36 & $1.08\times 10^6$ \\
C            & 370  & 106   & 29.5 & $1.90\times 10^6$ \\
\noalign{\smallskip}\hline
\end{tabular}
\caption{AMMT machine: experimental measurements according to~\cite{ambench2018}, CHAL-AMB2018-02-MP~ \label{tab:AMMTMeasurements}}
\end{table}

\subsection{Isotropic conductivity model} \label{sec:ModelCalibrationStandardHeatSource}
\begin{figure}[h]
     \includegraphics[width=0.45\textwidth]{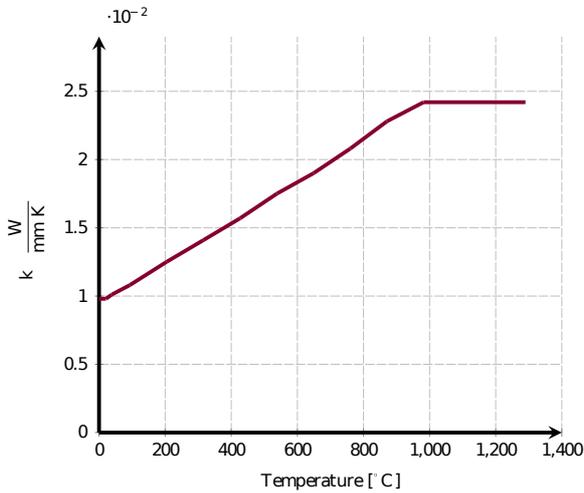}
		\caption{Conductivity vs. Temperature \label{fig:Cond} }
\end{figure}

\begin{figure}[h]
      \includegraphics[width=0.45\textwidth]{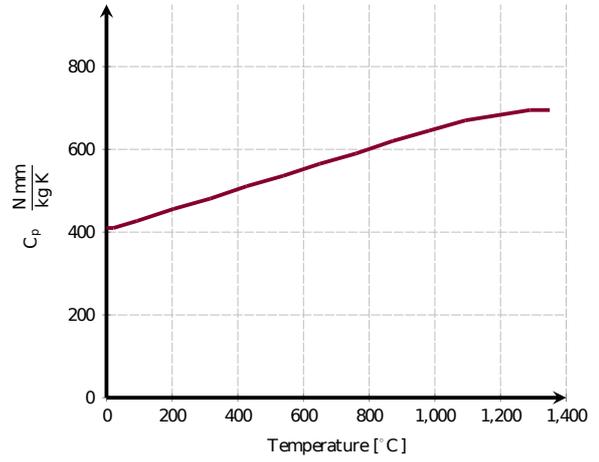} 
  \caption{Heat capacity vs. Temperature \label{fig:HeatCap} }
\end{figure}

The calibration step of the isotropic model is carried out for case B on the CBM machine,  as given in~\autoref{tab:CBSparameterValues}, which is exactly the same configuration as case 7 in~\cite{Ghosh2018}.
The validation step is obtained comparing the calibrated model to the cases A, B and C of~\autoref{tab:CBSMeasurements} and~\autoref{tab:AMMTMeasurements}.
For all the numerical simulations the IN625 material parameters are taken from literature~\cite{specialmetals,Mills2002} and are reported in~\autoref{materialConstPar},~\autoref{fig:Cond}, and \autoref{fig:HeatCap}.
\par
\begin{table}[h]
\centering
\caption{Material and process constant parameters}
\label{materialConstPar}
\begin{tabular}{|l|r|}
\hline
density   	   		 	         & 8.44e-6 {[}kg/mm$^3${]}   \\
\hline
latent heat            		     & 2.8e5 {[}J/kg{]}  \\
\hline
melting temperature interval     & 1290 - 1350 {[}$^{\circ}$C{]}  \\
\hline
\end{tabular}
\end{table}

It is noteworthy that material and process parameters, necessary to run the numerical simulation, are not experimentally available for the effective temperatures occurring in LPBF processes. 
For example, the measurement of the thermal conductivity $k$ in~\autoref{fig:Cond} is only available up to $871^\circ$C, but the melting range for IN625 is 1290-1350$^{\circ}$C. Likewise, the melting temperature interval the value of $k$ can only be extrapolated. It is important to note that this extrapolation itself represents a physical model which, in turn, needs to be calibrated. 
This circumstance is used in
~\autoref{sec:anisotrConductivity} to better describe the conductivity of the material and, consequently, improve the accuracy of the predicted melt pool geometry. 
Further coefficients, whose measurements are only available up to a certain temperature, are the absorptivity $\eta$ and the emissivity $\epsilon$. The latter necessary to define the radiation boundary condition given in~\autoref{eq:emissivity}.
\par
\begin{figure}[ht]
\centering
  \subfloat[Length]
	{
     \includegraphics[width=0.24\textwidth]{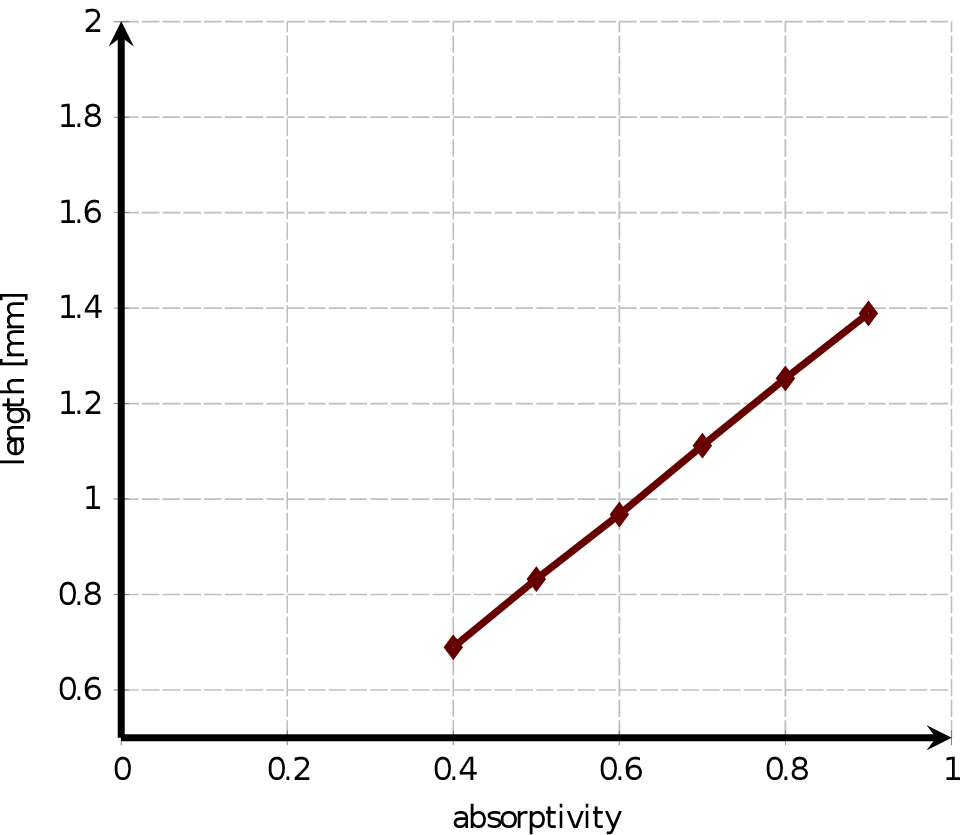}
  }
 \subfloat[Depth]
	{
     \includegraphics[width=0.24\textwidth]{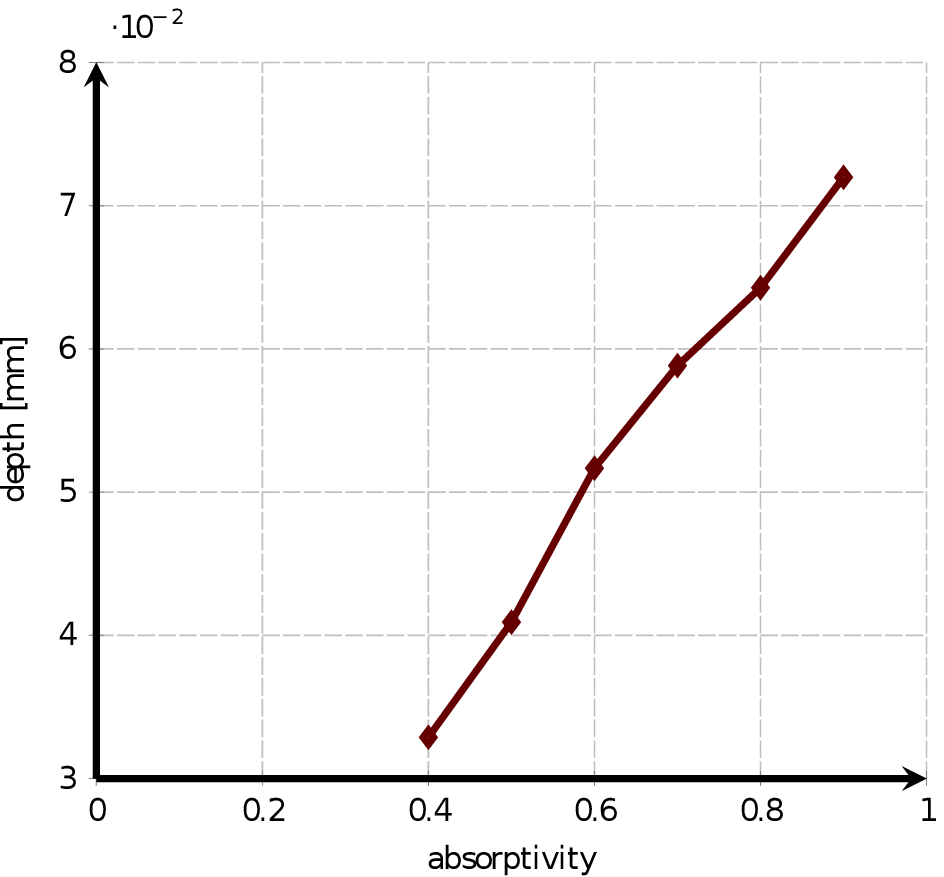}
  }
  \\
  \subfloat[Width]
	{
     \includegraphics[width=0.24\textwidth]{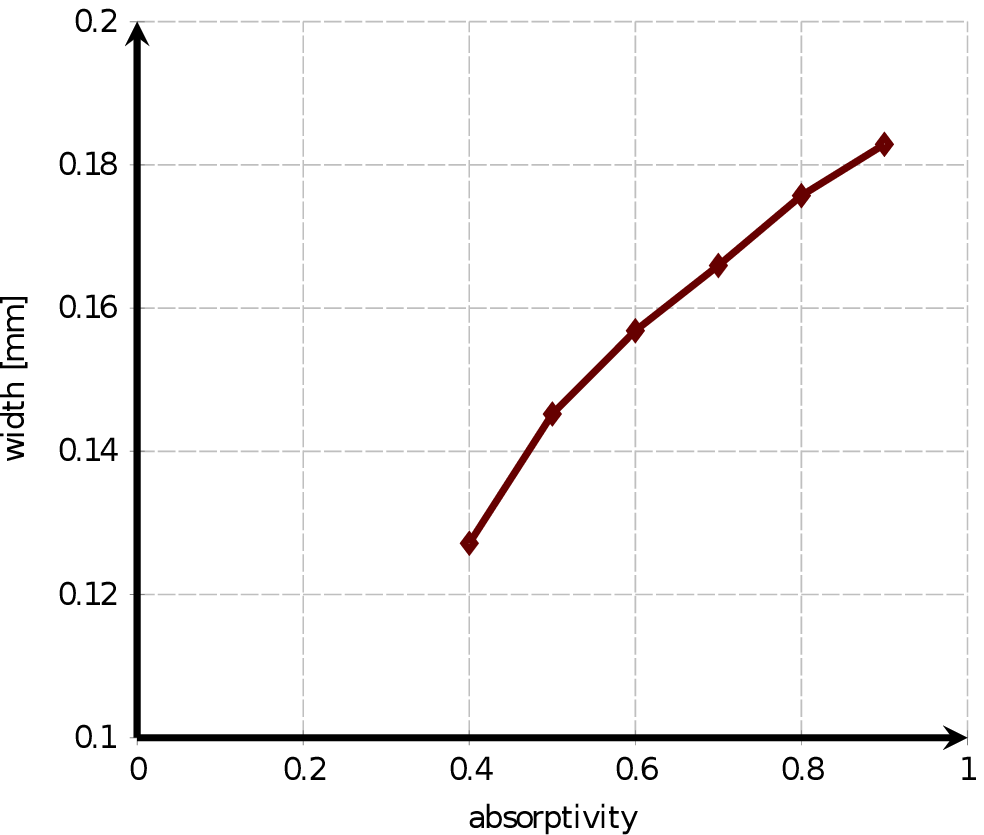}
  }
    \subfloat[Cooling rate]
	{
     \includegraphics[width=0.24\textwidth]{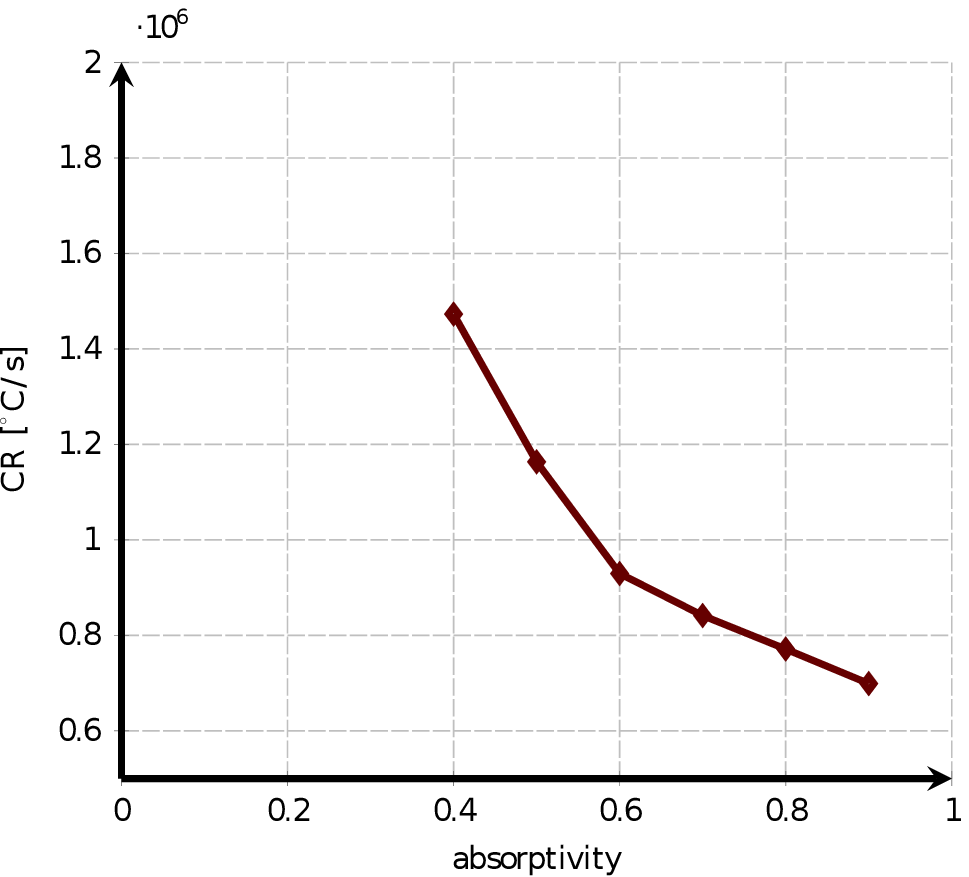}
  }

\caption{CBM machine. Sensitivity studies w.r.t. the absorptivity of the material $\eta$ \label{fig:AbsorptivityStudy}}
\end{figure}
\begin{figure}[bt]
     \includegraphics[width=0.45\textwidth]{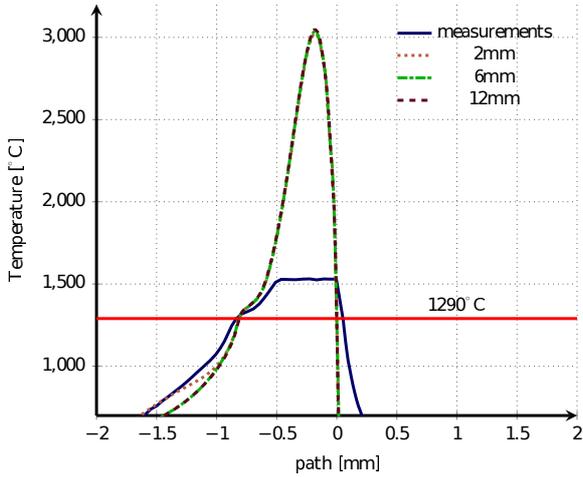}
       \caption{Computation of the temperature profile calibrated to case 7 in~\cite{Ghosh2018} \label{TemProfCal}}
\end{figure}
\begin{figure}[bt]
      \includegraphics[width=0.45\textwidth]{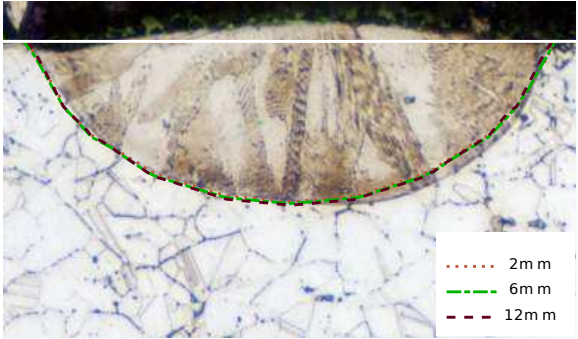} 
      \caption{Computation of cross section \label{CrossSecCal} (red line) calibrated to case 7 in~\cite{Ghosh2018}}
\end{figure}

In case of the CBM machine, the benchmark defines the laser spot radius equal to $50\mu m$. We utilize this value for both parameters $c_f$ and $a$ of the double elliptical model (see~\autoref{GaussDist}).
However, our model also contains the radius ratio $c_r/c_f$ as a model parameter as well as the power fraction $f_f/f_r$. Both these parameters are additional, potential candidates to calibrate the physical model. 
\paragraph{Model Calibration}
\mbox{}\\
Model calibration first requires to identify the sensitivities of the quantities of interest, i.e., length, width, depth, and cooling rate at the wake of the melt pool w.r.t. the modeling parameters $\eta$, $\epsilon$, $f_f/f_r$ and $c_r/c_f$ given by the physical model presented in~\autoref{sec:governingEquations}. 
To this end, four studies were carried out, and for each study a single  parameter is varied while the others stay fixed. In the following we only present the sensitivity studies w.r.t. the absorptivity $\eta$ and the emissivity $\epsilon$ of the material, while the other results are provided in the additional material of the present article. 
\begin{figure}[ht]
\centering
  \subfloat[Length]
	{
     \includegraphics[width=0.24\textwidth]{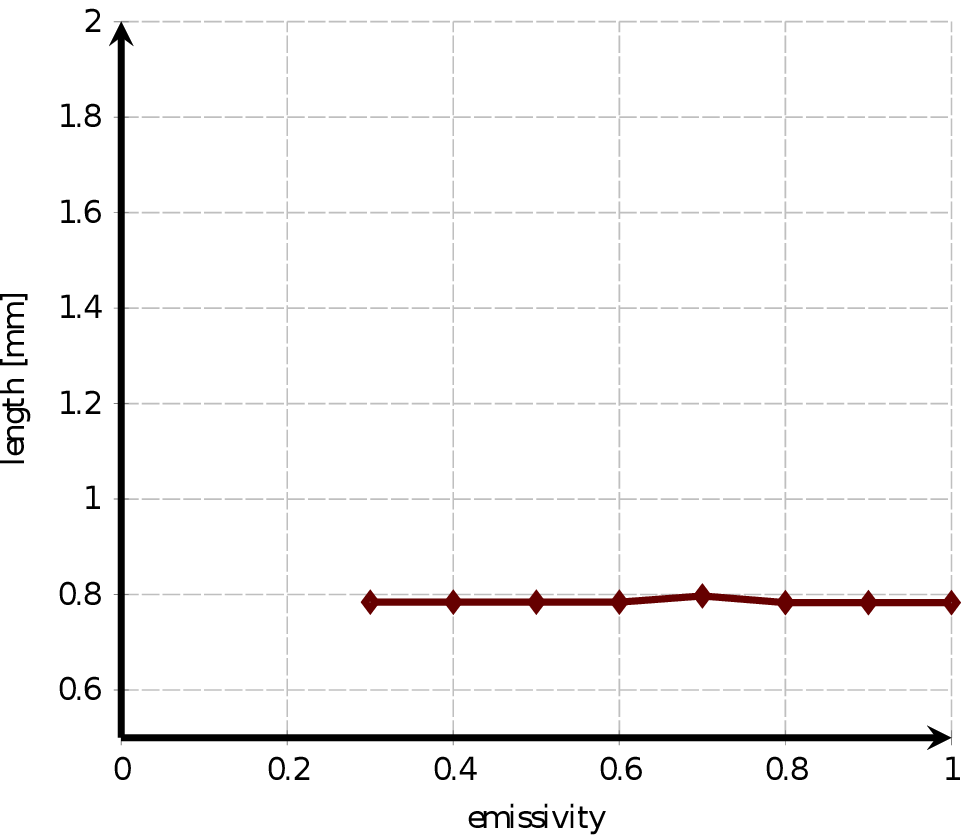}
  }
 \subfloat[Depth]
	{
     \includegraphics[width=0.24\textwidth]{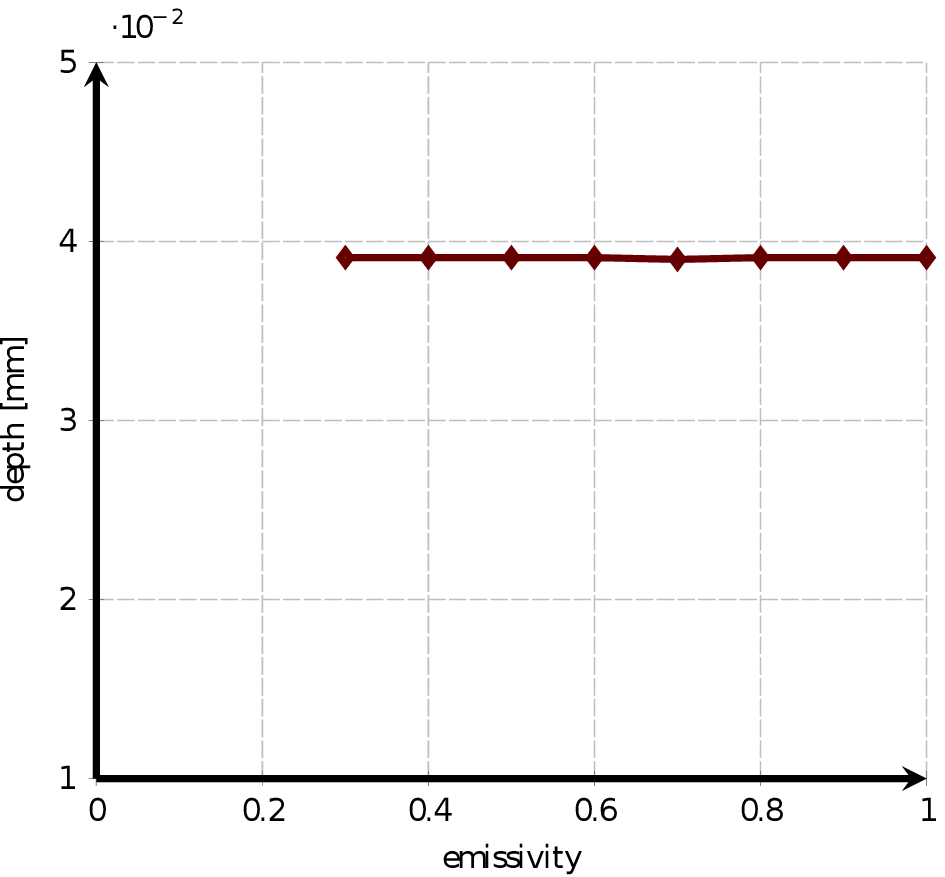}
  }
  \\
  \subfloat[Width]
	{
     \includegraphics[width=0.24\textwidth]{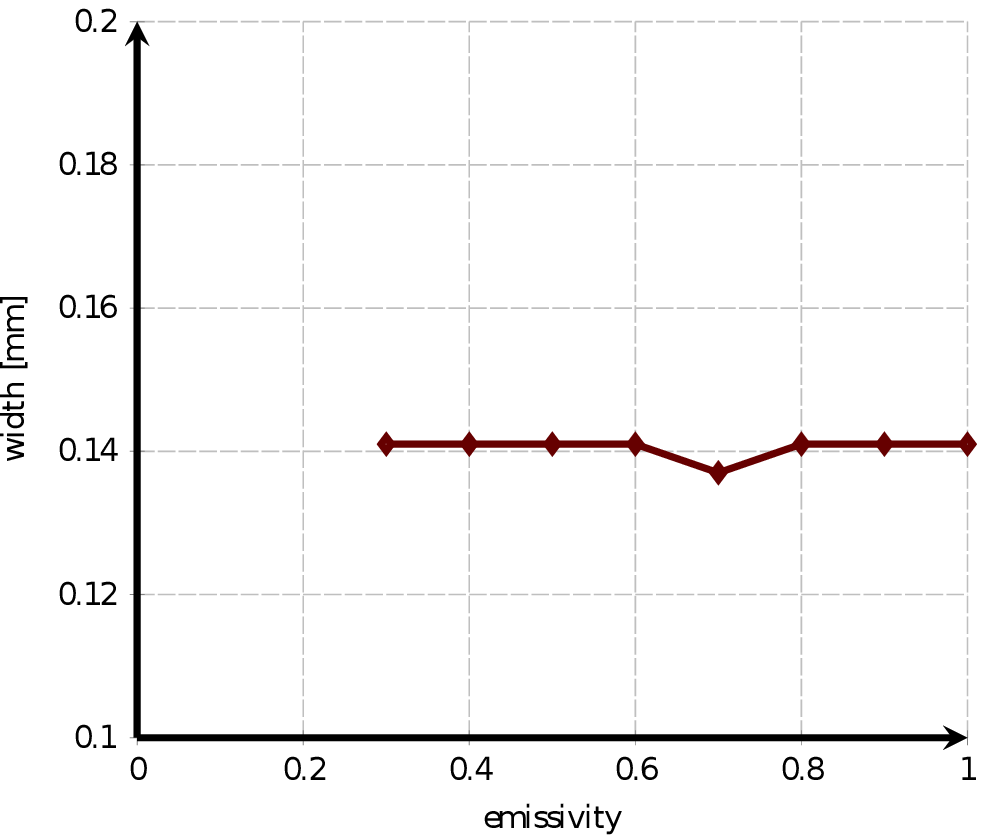}
  }
    \subfloat[Cooling rate]
	{
     \includegraphics[width=0.24\textwidth]{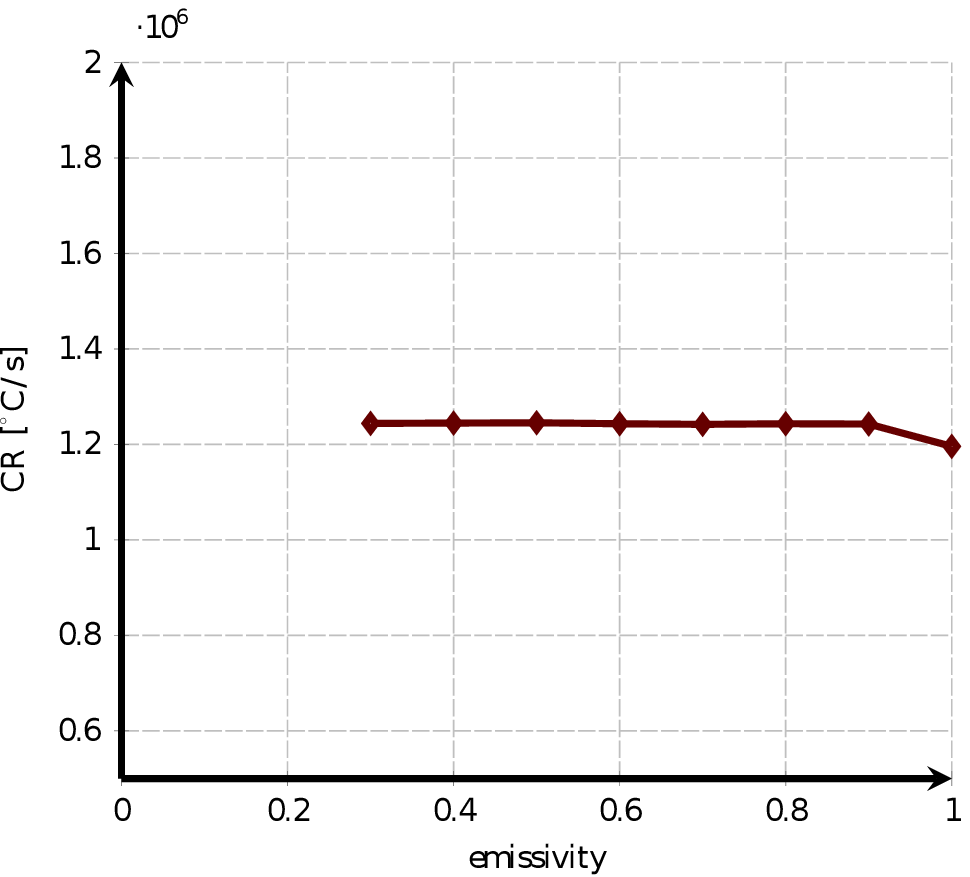}
  }

\caption{CBM machine. Sensitivity studies w.r.t. the emissivity of the material $\epsilon$ \label{fig:EmissivityStudy}}
\end{figure}
\autoref{fig:EmissivityStudy} presents the variation of the length, width, depth, and cooling rate w.r.t. the emissivity values.
The study clearly suggests that there is practically no influence of the emissivity on the quantities of interest. 
At first sight this result comes as a surprise because the model of the boundary conditions suggests an influence of fourth order in the temperature, something that for sure can not be neglected. 
Indeed, numerous authors explicitly include this boundary condition to obtain good results, see for example~\cite{Lu2018} and the references cited therein.  
However, the investigation at hand considers the temperature directly under or in close vicinity to the laser and, therefore, the contribution of the radiation boundary condition is marginal. 
To illustrate this effect, we consider the flux caused by radiation at the melting temperature $T_m=1290^\circ$C$=1563.15$K, with an ambient temperature of $T_e=20^\circ$C$=293.15$K. 
The corresponding power loss is 
\begin{multline}
5.67  \times 10^{-8}
\times 0.47
\left(T_m
^2+T_e
^2\right) 
\left( T_e
^2-T_m
^2\right) 
\\
= 1.59\times10^5 \left[\frac{\text{W}}{m^2}\right] 
= 0.16\left[\frac{\text{W}}{mm^2}\right],
\end{multline}
which represents a negligible quantity compared to the peak power density of $q^l=2.33\times10^4\left[W/mm^2\right]$ in the center of the laser beam. Clearly, under these conditions, radiation itself may be neglected for studies of temperature fields in close proximity to the laser source. 
To the contrary, the absorptivity has a large influence (see~\autoref{fig:AbsorptivityStudy}), as do the power fraction and the radius ratio\footnote{The latter two are not depicted due to limitations of space in this article}. 
An iterative calibration delivers the final choice of the parameters: $\varepsilon=0.47$, $\eta=0.38$, $f_f/f_r=0.053$ and $c_r/c_f=0.167$.
\paragraph{Numerical results for calibration}
\mbox{}\\
Numerically computed temperature curves along the laser path are depicted in~\autoref{TemProfCal}. The figure reports also the experimentally measured temperature, carried out using \textit{in-situ} thermography, as described in~\cite{Ghosh2018}. 
The different curves labeled $2mm$, $6mm$ and $12mm$ indicate at which position the zero of the abscissa of the plot coincides with the laser path. A steady state is reached already after only $2mm$. 
We specifically note that the calibration was carried out to best capture the temperature range around the melting temperature. 
Larger, even unphysical deviations, are tolerated outside this region.
This kind of calibration towards a process window is justifiable not only due to the fact that merely the region of interest needs to be captured with accuracy by the computations, but also because the camera itself delivers its most accurate measurements in that range. 
The plot also directly shows where the numerical model is not valid, namely directly inside the melt pool. 
Here, the temperature drastically overshoots to unrealistically high values. 
The very good agreement of the computation in the range of the melting zone is further confirmed in~\autoref{CrossSecCal}. 
This figure overlays the image of the cross section of the track taken by an ex-situ measurement of a confocal laser scanning microscope (CLSM) with the calibrated computation. 
Both ~\autoref{TemProfCal} and~\autoref{CrossSecCal} demonstrate that it is possible to obtain an excellent agreement with the experiment using the simple physical model presented in~\autoref{sec:governingEquations}, if $\eta$, $f_f/f_r$, and $c_r/c_f$ serve as model calibration parameters.
\paragraph{Model validation for the CBM machine}
\mbox{}\\
{
The calibrated model delivers the results depicted in~\autoref{tab:ResultsCBSCase}.
\begin{table}
\begin{tabular}{lllllll}
\hline\noalign{\smallskip}
case                  & meas.  & num.            & $\Delta$ & meas.            & num.            & $\Delta$ \\
                      & l$[\mu m]$ &  l$[\mu m]$       &  [\%]        &  CR[$\frac{^\circ\text{C}}{sec}$]   & CR [$\frac{^\circ\text{C}}{sec}$]  & [\%]  \\
\noalign{\smallskip}\hline\noalign{\smallskip}
A            & 659 & 707   & 7.3 & 6.20$\times 10^5$   & 8.79$\times 10^5$ & 41.8 \\
B             & 782 & 812   & 3.8 & 9.35$\times 10^5$   & 1.35$\times 10^6$ & 44.3\\
C            & 754  & 772   & 2.4  &  1.28$\times 10^6$ & 2.09$\times 10^6$ & 63.3 \\
\noalign{\smallskip}\hline
\end{tabular}
\caption{CBM machine: obtained weld pool length l and cooling rates cr \label{tab:ResultsCBSCase} }
\end{table}
It can readily be concluded that the model is able to predict the length of the weld pool up to at least 7.3\% accuracy in the parameter range covered by cases A to C. The prediction of cooling rates is approximately one order less accurate. In fact, the cooling rate is a derived variable (we directly compute only the temperature) and thus less accuracy is naturally expected; moreover, since our model does neither include convection nor evaporation, the temperature predicted within the melt pool is likely inaccurate and this surely affects also the cooling rate results in the transfer region from solid to liquid.

It is interesting to note that the measured length in~\cite{Ghosh2018}, i.e., the length towards which the model was calibrated was 813$[\mu m]$, provided with a tolerance of $\pm 79[\mu m]$. However, measurements performed in~\cite{ambench2018} for exactly the same case (case B) were more accurate and are given as $782\pm 21[\mu m]$, see also~\autoref{tab:ResultsCBSCase}. Thus, a re-calibration of the model to case B will likely deliver more accurate predictions for the cases A and C. However, this was not carried out because even more accurate measurements are available for the AMMT machine which lead to the development of the extended physical model presented in~\autoref{sec:anisotrConductivity}.
}
\paragraph{Model validation for the AMMT machine}
\mbox{}\\
{
Surprisingly, very different experimental results were obtained with the same scan parameters at the AMMT machine as compared to the CBM machine.
Due to this reason, more thorough studies were carried out on the AMMT machine. These include measurements of the actual laser profile itself. These measurements, now published in~\cite{ambench2018}, enable their direct application as the Neumann boundary condition $q^l$ in~\autoref{eq:emissivity}. Thus, the physical model presented in~\autoref{sec:governingEquations} is more tightly defined. 
This generates an interesting situation from the perspective of model validation because two (influential) calibration parameters, the power fraction $f_f/f_r$ and the radius ratio $c_r/c_f$ are now fixed and, therefore, can not be used for calibration. 
Given that the emissivity $\epsilon$ has practically no influence, the absorptivity $\eta$ is the only parameter left for a re-calibration. For $\eta=0.086$ we obtain the numerical results provided in~\autoref{tab:ResultsAMMTCaseOneParameterModel}. 
\begin{table}
\centering
\begin{tabular}{lllll}
\hline\noalign{\smallskip}
case                  & length    & width            & depth             & cooling rate             \\
                      & $[\mu m]$ &  $[\mu m]$       &  $[\mu m]$        &  [$\frac{^\circ\text{C}}{sec}$]   \\
\noalign{\smallskip}\hline\noalign{\smallskip}
A            & 301  & 119   & 52 & 0.91$\times 10^6$ \\
B             & 360 & 103   & 42 & 1.33$\times 10^6$ \\
C            & 348  & 91   & 32 & 2.18$\times 10^6$ \\
\noalign{\smallskip}\hline
\end{tabular}
\caption{AMMT machine: computed values \label{tab:ResultsAMMTCaseOneParameterModel}}
\end{table}
The corresponding deviations are provided in~\autoref{tab:DeviationsAMMTCaseOneParameterModel}.
\begin{table}
\centering
\begin{tabular}{lllll}
\hline\noalign{\smallskip}
case                  & length    & width            & depth             & cooling rate             \\
                      & $\Delta[\%]$ &  $\Delta[\%]$  &  $\Delta[\%]$    &  $\Delta[\%]$   \\
\noalign{\smallskip}\hline\noalign{\smallskip}
A            & 0.47  & 19.3   & 18.6 & 21.6 \\
B            & 0.11  & 16.4   & 15.8 & 23.1 \\
C            & 5.9   & 14.2   & 10.1 & 14.7 \\
\noalign{\smallskip}\hline
\end{tabular}
\caption{AMMT machine: deviations from experimental values \label{tab:DeviationsAMMTCaseOneParameterModel}}
\end{table}
While the deviations in the length are still at a maximum of approx. 6\%, width and depth are only predicted to an accuracy of 20\%. No further calibration is possible as there is only one parameter to calibrate but three values of interest to fit (excluding the cooling rate).
This clearly shows the boundaries of  validity of the model presented in~\autoref{sec:governingEquations} and motivates the development of the model discussed in the next section. 
}
%
\subsection{Anisotropic conductivity model} \label{sec:anisotrConductivity}
\begin{figure}
  \includegraphics[width=\columnwidth]{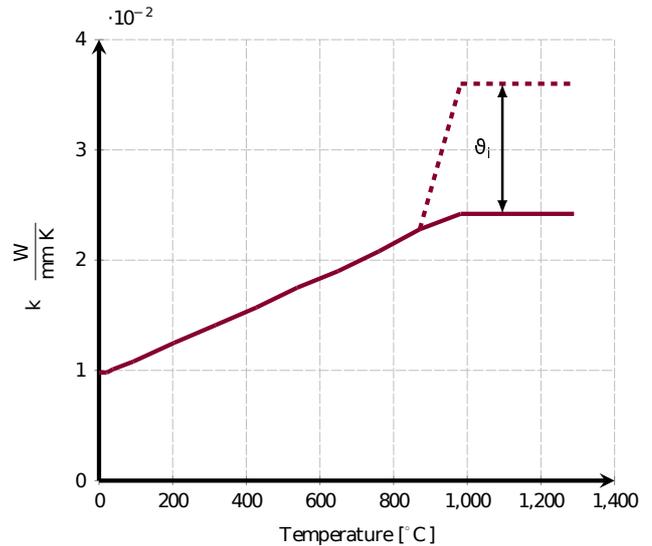}
\caption{Scaling of the conductivity in direction $i={x,y,z}$}
\label{fig:anisitropicConductivity}       
\end{figure}

\begin{figure*}
  \includegraphics[width=\textwidth]{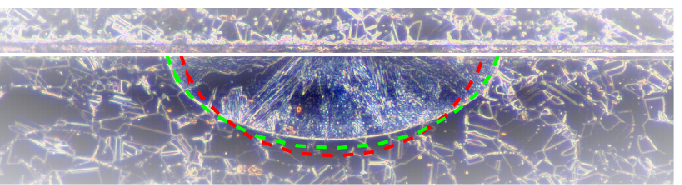}
\caption{Melt pool cross section micrograph image $50\times$DF (from \url{https://phasedata.nist.gov/rest/blob?id=5b102edd4407e700870ff13e}) over computed cross sections using isotropic (dashed red line) and anisotropic conductivity (dashed green line)}
\label{fig:crossSection}       
\end{figure*}

Two possible modifications are readily imaginable: a) a definition of an absorptivity field instead of a scalar value $\eta$ and b) the definition of an anisotropic conductivity. 
The former could be motivated by the fact that the melt pool surface will surely cause the absorption of the laser energy to be non-constant. 
However, to the authors opinion a good model should be as simple as possible, yet replicate the observed effects as accurately as possible. 
With this objective in mind, the definition of an anisotropic conductivity is a more attractive choice.
The only change necessary is that the scalar value $k$ in \autoref{eq:transientHeat} changes to $\mathbf{k}$, a diagonal matrix with the entries $diag(k_x$, $k_y$, $k_z$). 
Further, we set $\epsilon=0$. 
The physical motivation for this model is that the (transient) diffusion equation given by~\autoref{eq:transientHeat} does by no means include the effects caused by convective heat transfer inside the weld pool. 
This flaw has already inspired other authors e.g.,~\cite{Lu2018} to use a strongly increased conductivity $k$ inside the melt pool to model convective effects. 
We now extend this idea by choosing anisotropic values. 
For simplicity, we introduce the scaling factor $\vartheta_i$ where $i=\{x,y,z\}$ such that $\mathbf{k}=diag(k\vartheta_x,k\vartheta_y,k\vartheta_z)$. 
The values for $\vartheta_i$ deviate from $1$ only after the last obtainable measurement (at $T=871^{\circ}C$) of the conductivity as depicted in~\autoref{fig:anisitropicConductivity}.
After calibration to the AMMT machine B we obtain the set $\vartheta_x=1.0,\vartheta_y=1.4, \vartheta_z=0.9$. 
This delivers very well matching weld pool geometries. 
While the effect of the scaling of $k$ is marginal in a  temperature plot along the length (because here $\vartheta_x=1.0$), its effect in the cross-section is quite pronounced (see~\autoref{fig:crossSection} for a direct overlay of the melt pool geometry over the cross section). 
In the validation step, we keep the calibration parameters fixed, i.e., $\vartheta_x=1.0$, $\vartheta_y=1.4$, and $\vartheta_z=0.9$ and we compute cases A and C of the AMMT machine. 
The computed values are provided in~\autoref{tab:ResultsAMMTCaseAnisotropicParameterModel} and the corresponding deviations are provided in~\autoref{tab:DeviationsAMMTCaseAnisotropicParameterModel}.
\begin{table}
\centering
\begin{tabular}{lllll}
\hline\noalign{\smallskip}
case                  & length    & width            & depth             & cooling rate             \\
                      & $[\mu m]$ &  $[\mu m]$       &  $[\mu m]$        &  [$\frac{^\circ\text{C}}{sec}$]   \\
\noalign{\smallskip}\hline\noalign{\smallskip}
A            &  304  & 146.4 & 44.6  & 0.82$\times 10^6$  \\
B             & 362  & 123.7 & 36.1  & 1.23$\times 10^6$  \\
C            &  346  & 105.1 & 27.3  & 1.88$\times 10^6$ \\
\noalign{\smallskip}\hline
\end{tabular}
\caption{Anisotropic conductivity model: computed values \label{tab:ResultsAMMTCaseAnisotropicParameterModel}}
\end{table}
\begin{table}
\centering
\begin{tabular}{lllll}
\hline\noalign{\smallskip}
case                  & length    & width            & depth             & cooling rate             \\
                      & $\Delta[\%]$ &  $\Delta[\%]$  &  $\Delta[\%]$    &  $\Delta[\%]$   \\
\noalign{\smallskip}\hline\noalign{\smallskip}
A            & 1.33  &  1.0  & 2.5 & 29.3 \\
B            & 0.84  &  0.02 & 0.2 & 13.9 \\
C            & 6.49  &  0.8  & 5.1 &  1.3\\
\noalign{\smallskip}\hline
\end{tabular}
\caption{Anisotropic conductivity model: deviations of computed values from experimental values \label{tab:DeviationsAMMTCaseAnisotropicParameterModel}}
\end{table}
We observe that for the anisotropic model the maximum deviation of length, width and depth is 6.49\% at worst while, for the isotropic conductivity model it was merely 19.3\%. 
Even the forecast of the cooling rates has improved sightly.
\section{Summary and Conclusions} \label{sec:Conclusions}

In this contribution we used the standard heat diffusion model to predict the length, width and depth of the melt pool in the laser additive manufacturing benchmarks CHAL-AMB2018-02-MP published in~\cite{ambench2018}. The physical model included a latent heat term as published e.g., in~\cite{Celentano1994} along with a radiation boundary condition. Within this model we found the radiation boundary condition to have little to no influence upon the quantities of interest. This is due to the fact that in close proximity of the laser beam impact region, the power lost by radiation is much lower than the applied laser energy itself. 

As a first approach we assumed the laser source to possess the well known double elliptical shape as proposed for welding by Goldak~\cite{Goldak1984}. We demonstrated that this model is well suited to predict the shape of the weld pool as it delivered a maximum deviation from the measurements of 7.3\%. However, in case the shape of the laser source is given by a measurement, the standard, transient heat diffusion model only provides accuracies of 19.1\% for the investigated benchmark cases. This renders it practically invalid.

We then extended the isotropic thermal model by introducing anisotropic conductivities. Their physical interpretation is to model anisotropic convection inside the melt pool. This slight extension enabled the model to deliver at worst 6.49\% deviations in length, width and depth of the melt pool. Therefore, we conclude that the introduction of an anisotropic conductivity is a simple, yet effective way to improve the physical model based on transient heat equation including phase changes and remark that the added computational effort for this extension is marginal.

In the future, we aim at validating a numerical model including powder. Nevertheless, this task is not straightforward since measurements of melt pool shapes in presence of a powder bed are extremely challenging. We want to finally stress the fact that the presented model is meant to provide, once thoroughly calibrated, a reliable model for multi-track and eventually multi-layer thermal simulations. However, it is not valid to predict the temperature distribution within the melt pool.

\section{Acknowledgements}
The first author gratefully acknowledges the financial support of the German Research Foundation (DFG) under grant RA 624/27-2.
This work was partially supported by Regione Lombardia through the project "TPro.SL - Tech Profiles for Smart Living" (No. 379384) within the Smart Living program, and through the project "MADE4LO - Metal ADditivE for LOmbardy" (No. 240963) within the POR FESR 2014-2020 program.
Massimo Carraturo and Alessandro Reali have been partially supported by Fondazione Cariplo - Regione Lombardia through the project ``Verso nuovi strumenti di simulazione super veloci ed accurati basati sull'analisi isogeometrica'', within the program RST - rafforzamento.

\bibliographystyle{spmpsci}      
\bibliography{SimulationInAppliedMechanicsatCiE}   

%
%
\end{document}